\documentclass[10pt,pra,twocolumn,showpacs,floatfix]{revtex4}
\usepackage{amsfonts}
\usepackage{amssymb}
\usepackage{amsmath}
\usepackage[dvips]{graphicx}

\begin{document}

\title{Five-wave-packet quantum error correction based on
continuous-variable cluster entanglement}
\author{Shuhong Hao, Xiaolong Su,}
\email{suxl@sxu.edu.cn}
\author{Caixing Tian, Changde Xie and Kunchi Peng}
\affiliation{State Key Laboratory of Quantum Optics and Quantum Optics Devices, \\
Institute of Opto-Electronics, Shanxi University, Taiyuan, 030006, People's
Republic of China}

\begin{abstract}
Quantum error correction protects the quantum state against noise and
decoherence in quantum communication and quantum computation, which enables
one to perform fault-torrent quantum information processing. We
experimentally demonstrate a quantum error correction scheme with a
five-wave-packet code against a single stochastic error, the original
theoretical model of which is firstly proposed by S. L. Braunstein and T. A.
Walker. Five submodes of a continuous variable cluster entangled state of
light are used for five encoding channels. Especially, in our encoding
scheme the information of the input state is only distributed on three of
the five channels and thus any error appearing in the remained two channels
never affects the output state, i.e. the output quantum state is immune from
the error in the two channels. The stochastic error on a single channel is
corrected for both vacuum and squeezed input states and the achieved
fidelities of the output states are beyond the corresponding classical limit.
\end{abstract}

\maketitle

\section{Introduction}

The transmission of quantum states with high fidelity is an essential
requirement for implementing quantum information processing with high
quality. However, losses and noises in channels inevitably lead to errors
into transmitted quantum states and thus make the distortion of resultant
states. The aim of quantum error correction (QEC) is to eliminate or, at
least, reduce the hazards resulting from the imperfect channels and to
ensure transmission of quantum states with high fidelity \cite{Nielsen2000}.
A variety of discrete variable QEC protocols, such as nine-qubit code \cite%
{Shor}, five-qubit code \cite{Laf1996}, topological code \cite%
{Dennis2001,Fowler2012}, have been suggested and the experiments of QEC have
been realized in different physical systems, such as nuclear magnetic
resonance \cite{Cory1998,Knill2001,Boulant2005}, ionic \cite%
{Chiaverini2004,Schindler2011}, photonic \cite{Yao2012,Bell2014},
superconducting systems \cite{Reed2012,Barends2014} and Rydberg atoms \cite%
{Ottaviani}.

Besides quantum information with discrete variables, quantum information
with continuous variables (CV) is also promptly developing \cite%
{RMP1,RMP2,Furusawa1998,Li2002,Menicucci2006,Gu2009,Ukai2011,Su2013}.
Different types of CV QEC codes for correcting single non-Gaussian error
have been proposed, such as nine-wave-packet code \cite%
{Braunstein1998,Lloyd1998}, five-wave-packet code \cite%
{Braunstein19982,Walker2010}, entanglement-assisted code \cite{Wilde2007}
and erasure-correcting code \cite{Niset2008}. A CV QEC scheme against
Gaussian noise with a non-Gaussian operation of photon counting has been
also theoretically analyzed \cite{Ralph2011}. The CV QEC schemes of the
nine-wave-packet code \cite{Aoki2009}, erasure-correcting code against
photon loss \cite{Lassen2010} and the correcting code with the correlated
noisy channels \cite{Lassen2013} have been experimentally demonstrated.

According to the no-go theorem proved in Ref. [34], Gaussian errors are
impossible to be corrected with pure Gaussian operations. However,
non-Gaussian stochastic errors, which frequently occur in free-space
channels with atmospheric fluctuations for example \cite%
{Heersink2006,Dong2008,Hage2008}, can be corrected by Gaussian schemes since
the no-go theorem does not apply in this case. Generally, the stochastic
error model is described by \cite{Loock2010} 
\begin{equation}
W_{out}(x,p)=(1-\gamma )W_{in}(x,p)+\gamma W_{error}(x,p),
\end{equation}%
where the input state $W_{in}(x,p)$\ is transformed into a new state $%
W_{error}(x,p)$\ with probability $\gamma $\ or it remains unchanged with
probability $1-\gamma $. Even for the case of two Gaussian states $%
W_{in}(x,p)$\ and $W_{error}(x,p)$, the output state $W_{out}(x,p)$\ is also
non-Gaussian, that is, this channel model describes a certain, simple form
of non-Gaussian errors.

In 2009 T. Aoki et al. presented the first experimental implementation of a
Shor-type nine-channel QEC code based on entanglement among nine optical
beams, which was the achievable largest entangled state on experiments then 
\cite{Aoki2009}. This scheme is deterministically implemented using only
linear operations and resources, which can correct arbitrary single beam
error. Although S. L. Braunstein discovered a highly efficient
five-wave-packet code theoretically in 1998, its linear optical construction
was not proposed \cite{Braunstein19982}. Later, in 2010, T. A. Walker and S.
L. Braunstein outlined a new approach for generating linear optics circuits
that encode QEC code and proposed a linear optics construction for a
five-wave-packet QEC code \cite{Walker2010}. Differentiating from previous
approaches by means of directly transferring existing qubit codes into CV
codes, they defined the conditions for yielding a CV QEC code firstly and
then searched numerically for circuits satisfying this criterion. The
five-wave-packet code improves on the capacity of the best known code
implemented by linear optics and saturates the lower bound for the number of
carrier needed for a single-error-correct code \cite{Walker2010}. However,
the proposed five-wave-packet CV QEC code has not been experimentally
demonstrated so far.

\begin{figure}[tbp]
\begin{center}
\includegraphics[width=80mm]{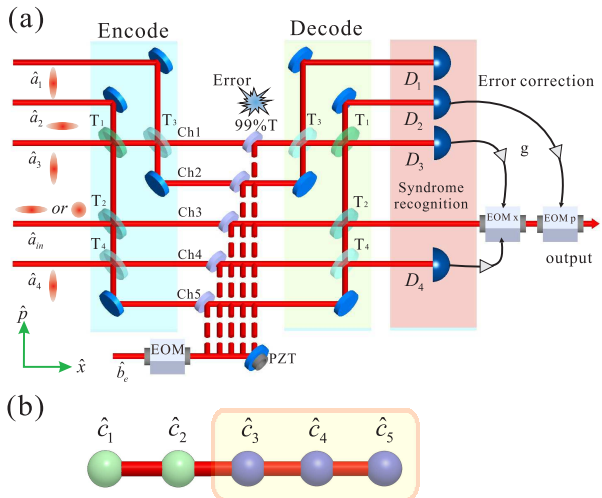}
\end{center}
\caption{(color online) The schematic of the QEC scheme. (a): the schematic
of experimental set-up. PZT: piezoelectric transducer. EOM: electro-optical
modulator, T$_{1-4}$: beam-splitters with 25\%, 33\%, 50\%, and 50\%
transmission, respectively. Ch1-5: quantum channels. 99\%T: a beam-splitter
with 99\% transmission. D$_{1}$-D$_{4}$: homodyne detectors, g: gain in the
feedforward circuit. (b): the graph representation of the five-wave-packet
code. The input state is encoded on submodes $\hat{c}_{3}$, $\hat{c}_{4}$
and $\hat{c}_{5}$ of a five-partite linear cluster state $\hat{c}_{1-5}$.}
\end{figure}

Based on the approach outlined by T. A. Walker and S. L. Braunstein \cite%
{Walker2010}, we design a more compact linear optics construction and
achieve the first experimental demonstration of five-wave-packet CV QEC code
using a five-partite CV cluster entangled state \cite{Zhang2006,Loock2007}.\
In this experiment only four ancilla squeezed states of light are required
and four optical beamsplitters are used in the encoding and the decoding
system, respectively. Comparing with the nine-wave-packet system \cite%
{Aoki2009}, the required quantum resources and utilized optical elements in
our system decrease a half. The smaller codes not only save quantum
resources, but also increase data rates and\textbf{\ }decrease the chance of
further occurring errors, thus are very significant for the development of
quantum information technology \cite{Walker2010}. In the presented encoding
method, only a part of all wave packets (three of five in the presented
experiment) involves the information of the input state and therefore the
noise occurring in the remained channels (channels 1 and 2 in the presented
system) do not introduce any error into the transmitted quantum state. Such
that, we do not need to perform the error correction for the remained
channels and the near unity fidelity is achieved in these channels. We name
the encoding method as the partial encoding. It should be emphasized that
although the remained two channels do not involve the information of the
input state, they play the unabsolvable roles in the syndrome recognition
and the error correction. In the presented QEC experiment, the error
correction is implemented in a deterministic fashion due to the application
of unconditional CV quantum entanglement \cite{RMP1,RMP2}. A vacuum state
and a squeezed vacuum state are utilized as the input states, respectively,
to exhibit the QEC ability of the system for different input states.
According to the standard notation for QEC code \cite{Nielsen2000}, the
presented five-wave-packet code should be expressed by $[n,k,d]=[5,1,3]$,
where $n=5$\ denote the number of used wave packets, $k=1$\ is the number of
logical encoded input state, and $d=3$\ is the distance, which indicates how
many errors can be tolerated, a code of distance $d$\ can correct up to $%
(d-1)/2$\ arbitrary errors at unspecified channels.

\section{Results}

\subsection{Encoding.}

The schematic of the CV QEC scheme is shown in Figure 1(a). The QEC
procedure contains five stages, which are encoding, error-in, decoding,
syndrome recognition and correction, respectively. The encoding is completed
by a beam-splitter network consisting of four beam-splitters (T$_{1}$-T$_{4}$%
). Four squeezed states with $-3.5$ dB squeezing $(\hat{a}_{1-4})$ generated
by three non-degenerate optical parametric amplifiers, are used as ancilla
modes (see APPENDIX A for details). In the experiment, three
amplitude-squeezed states, $\hat{a}_{m}=e^{-r}\hat{x}_{m}^{(0)}+ie^{+r}\hat{p%
}_{m}^{(0)}$ $(m=1,3,4)$, and a phase-squeezed state, $\hat{a}_{n}=e^{+r}%
\hat{x}_{n}^{(0)}+ie^{-r}\hat{p}_{n}^{(0)}$ $(n=2)$ are applied, where $r$
is the squeezing parameter ($r=0$ and $r=+\infty $ correspond to no
squeezing and perfect squeezing, respectively), $\hat{x}_{j}^{(0)}$\ and $%
\hat{p}_{j}^{(0)}$ denote the amplitude and phase quadratures of the vacuum
field, respectively. The transformation matrix of the encoding network is
expressed by%
\begin{equation}
U=%
\begin{pmatrix}
\frac{1}{\sqrt{2}} & \frac{\sqrt{3}}{2\sqrt{2}} & \frac{1}{2\sqrt{2}} & 0 & 0
\\ 
\frac{1}{\sqrt{2}} & \frac{-\sqrt{3}}{2\sqrt{2}} & \frac{-1}{2\sqrt{2}} & 0
& 0 \\ 
0 & \frac{1}{\sqrt{6}} & \frac{-1}{\sqrt{2}} & \frac{1}{\sqrt{3}} & 0 \\ 
0 & \frac{1}{2\sqrt{6}} & \frac{-1}{2\sqrt{2}} & \frac{-1}{\sqrt{3}} & \frac{%
1}{\sqrt{2}} \\ 
0 & \frac{-1}{2\sqrt{6}} & \frac{1}{2\sqrt{2}} & \frac{1}{\sqrt{3}} & \frac{1%
}{\sqrt{2}}%
\end{pmatrix}%
.
\end{equation}%
The unitary matrix can be decomposed by $%
U=B_{45}^{-}(1/2)B_{34}^{+}(1/3)B_{12}^{+}(1/2)B_{23}^{+}(1/4)$. Here, $%
B_{kl}^{\pm }(T)$ stands for the transformation of modes $k$ and $l$ on a
beam-splitter,\textbf{\ }the corresponding transformation matrix is given by 
\begin{equation}
B^{\pm }=%
\begin{pmatrix}
\sqrt{1-T} & \sqrt{T} \\ 
\pm \sqrt{T} & \mp \sqrt{1-T}%
\end{pmatrix}%
.
\end{equation}

The input state $\hat{a}_{in}$\ is encoded with the four ancilla modes by $(%
\hat{c}_{1},\hat{c}_{2},\hat{c}_{3},\hat{c}_{4},\hat{c}_{5})^{T}=U(\hat{a}%
_{1},\hat{a}_{2},\hat{a}_{3},\hat{a}_{in},\hat{a}_{4})^{T}$. The encoded
five modes are 
\begin{eqnarray}
\hat{c}_{1} &=&\frac{\hat{a}_{1}}{\sqrt{2}}+\frac{\sqrt{3}\hat{a}_{2}}{2%
\sqrt{2}}+\frac{\hat{a}_{3}}{2\sqrt{2}},  \notag \\
\hat{c}_{2} &=&\frac{\hat{a}_{1}}{\sqrt{2}}-\frac{\sqrt{3}\hat{a}_{2}}{2%
\sqrt{2}}-\frac{\hat{a}_{3}}{2\sqrt{2}},  \notag \\
\hat{c}_{3} &=&\frac{\hat{a}_{2}}{\sqrt{6}}-\frac{\hat{a}_{3}}{\sqrt{2}}+%
\frac{\hat{a}_{in}}{\sqrt{3}},  \notag \\
\hat{c}_{4} &=&\frac{\hat{a}_{2}}{2\sqrt{6}}-\frac{\hat{a}_{3}}{2\sqrt{2}}+%
\frac{\hat{a}_{4}}{\sqrt{2}}-\frac{\hat{a}_{in}}{\sqrt{3}},  \notag \\
\hat{c}_{5} &=&\frac{-\hat{a}_{2}}{2\sqrt{6}}+\frac{\hat{a}_{3}}{2\sqrt{2}}+%
\frac{\hat{a}_{4}}{\sqrt{2}}+\frac{\hat{a}_{in}}{\sqrt{3}}.  \label{encode}
\end{eqnarray}%
From equation (\ref{encode}) we can see, the input state is partially
encoded on channels 3, 4 and 5 ($\hat{c}_{3}$, $\hat{c}_{4}$ and $\hat{c}%
_{5} $) by means of the designed beam-splitter network, while the encoded
states in channels 1 and 2 ($\hat{c}_{1}$ and $\hat{c}_{2}$) do not contain
any information of the input state.

As shown in Figure 1(b) the encoded five modes $\hat{c}_{i}$\textbf{\ (}$%
i=1,...,5$) is the five submodes of a five-partite CV linear cluster
entangled state \cite{Zhang2006,Loock2007}. The correlation noises of
quadrature components among the encoded five wave-packets are expressed by $%
\hat{x}_{c1}+\hat{x}_{c2}=\sqrt{2}\hat{x}_{1}^{(0)}e^{-r}$, $\hat{p}_{c2}-%
\hat{p}_{c1}-\hat{p}_{c3}=(-2\sqrt{2}\hat{p}_{2}^{(0)}e^{-r}-\hat{p}_{in})/%
\sqrt{3}$, $\hat{x}_{c3}+\hat{x}_{c2}+\hat{x}_{c4}=(\hat{x}_{1}^{(0)}e^{-r}-2%
\hat{x}_{3}^{(0)}e^{-r}+\hat{x}_{4}^{(0)}e^{-r})/\sqrt{2}$, $\hat{p}_{c4}-%
\hat{p}_{c3}-\hat{p}_{c5}=-\sqrt{3}\hat{p}_{in}$, and $\hat{x}_{c4}+\hat{x}%
_{c5}=\sqrt{2}\hat{x}_{4}^{(0)}e^{-r}$. These expressions show that the
correlation noises of $\hat{x}_{c1}+\hat{x}_{c2}$, $\hat{x}_{c3}+\hat{x}%
_{c2}+\hat{x}_{c4}$ and $\hat{x}_{c4}+\hat{x}_{c5}$ are smaller than the
corresponding normalized shot-noise-level (SNL) for any non-zero squeezing
of the ancilla modes. While the correlation noises of $\hat{p}_{c2}-\hat{p}%
_{c1}-\hat{p}_{c3}$\ and $\hat{p}_{c4}-\hat{p}_{c3}-\hat{p}_{c5}$\ depend on
the input state, i.e. they have different values for different input state.
The inseparability criteria of the five-mode cluster entangled state are
denoted by \cite{Loock2003}%
\begin{eqnarray}
\left\langle \Delta (\hat{x}_{c1}+\hat{x}_{c2})^{2}\right\rangle
+\left\langle \Delta (\hat{p}_{c2}-\hat{p}_{c1}-g_{3}\hat{p}%
_{c3})^{2}\right\rangle &<&1,  \notag \\
\left\langle \Delta (\hat{p}_{c2}-g_{1}\hat{p}_{c1}-\hat{p}%
_{c3})^{2}\right\rangle +\left\langle \Delta (\hat{x}_{c3}+\hat{x}_{c2}+g_{4}%
\hat{x}_{c4})^{2}\right\rangle &<&1,  \notag \\
\left\langle \Delta (\hat{x}_{c3}+g_{2}\hat{x}_{c2}+\hat{x}%
_{c4})^{2}\right\rangle +\left\langle \Delta (\hat{p}_{c4}-\hat{p}_{c3}-g_{5}%
\hat{p}_{c5})^{2}\right\rangle &<&1,  \notag \\
\left\langle \Delta (\hat{p}_{c4}-g_{6}\hat{p}_{c3}-\hat{p}%
_{c5})^{2}\right\rangle +\left\langle \Delta (\hat{x}_{c4}+\hat{x}%
_{c5})^{2}\right\rangle &<&1.  \notag \\
&&
\end{eqnarray}%
When all combinations of correlation variances on the left of the
inequalities (5) are less than the normalized boundary on the right side,
the five-wave-packet optical state is a CV cluster entangled state. With a
vacuum input state and choosing the optimal gains of $g_{i}\quad (i=1,2...6)$
the inseparability criteria will be satisfied for any non-zero squeezing of
the ancilla modes. In this case, the encoded five wave packets form a
five-partite linear cluster entangled state.

\subsection{Error-in.}

The five encoded wave packets constitute five quantum channels, where the
errors possibly occur. In the experiment, the noise is modulated on an
excess optical beam ($\hat{b}_{e}$) by an electro-optical modulator (EOM)
drove by a sin-wave signal at 2 MHz to make an error beam firstly. Then, the
error beam is randomly coupled into any one of the five coded wave packets
each time by a mirror of 99\% transmission. By sweeping the phase of the
error wave packet with the piezoelectric translator (PZT) attached on a
reflection mirror, a quasi-random displacement error is added on one of the
five channels. The experimental operation corresponds to adding an error
operator $\hat{e}_{i}\quad (i=1,2...5)$ on a corresponding optical wave
packet, the mathematic expression of which is $U(\hat{a}_{1},\hat{a}_{2},%
\hat{a}_{3},\hat{a}_{in},\hat{a}_{4})^{T}+(\hat{e}_{1},\hat{e}_{2},\hat{e}%
_{3},\hat{e}_{4},\hat{e}_{5})^{T}$, where only one of $\hat{e}_{i}$\ is
non-zero when an error is occurring in one channel.

\subsection{Decoding.}

The decoding circuit is the inverse of the encoding circuit. After decoding,
the output mode ($\hat{d}_{out}$) and syndrome modes ($\hat{d}_{1}$, $\hat{d}%
_{2} $, $\hat{d}_{3}$ and $\hat{d}_{4}$) of the five channels are calculated
by $U^{-1}[U(\hat{a}_{1},\hat{a}_{2},\hat{a}_{3},\hat{a}_{in},\hat{a}%
_{4})^{T}+(\hat{e}_{1},\hat{e}_{2},\hat{e}_{3},\hat{e}_{4},\hat{e}_{5})^{T}]$%
. The decoded modes are 
\begin{eqnarray}
\hat{d}_{1} &=&\hat{a}_{1}+\frac{\hat{e}_{1}+\hat{e}_{2}}{\sqrt{2}},  \notag
\\
\hat{d}_{2} &=&\hat{a}_{2}+\frac{3\hat{e}_{1}-3\hat{e}_{2}+2\hat{e}_{3}+\hat{%
e}_{4}-\hat{e}_{5}}{2\sqrt{6}},  \notag \\
\hat{d}_{3} &=&\hat{a}_{3}+\frac{\hat{e}_{1}-\hat{e}_{2}-2\hat{e}_{3}-\hat{e}%
_{4}+\hat{e}_{5}}{2\sqrt{2}},  \notag \\
\hat{d}_{out} &=&\hat{a}_{in}+\frac{\hat{e}_{3}-\hat{e}_{4}+\hat{e}_{5}}{%
\sqrt{3}},  \notag \\
\hat{d}_{4} &=&\hat{a}_{4}+\frac{\hat{e}_{4}+\hat{e}_{5}}{\sqrt{2}}.
\end{eqnarray}%
It is obvious that the input state and ancilla modes are recovered after the
decoding stage and the errors are included in five output channels. Please
note that the output state $\hat{d}_{out}$ does not contain the errors $\hat{%
e}_{1}$ and $\hat{e}_{2}$, which means that the output state is immune from
errors in channels 1 and 2. If the error occurs in channels 1 and 2, the
output state will not be affected.

\begin{figure*}[tbp]
\begin{center}
\includegraphics[width=120mm]{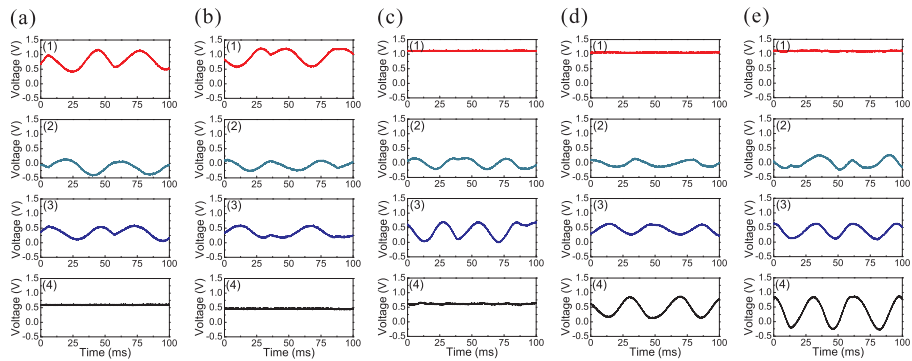}
\end{center}
\caption{(color online) Error syndrome measurement results. (a)-(e)
correspond to that a random displacement error is imposed on channel 1-5,
respectively. The DC outputs of detectors D1-D4 are recorded by a
four-channel digital oscilloscope and the results are shown in (1)-(4) from
top to bottom, respectively.}
\end{figure*}

\subsection{Syndrome measurement.}

From the decoded modes, we can see that the error in different channels
results in different outputs of the homodyne detectors D$_{1}$-D$_{4}$. By
the DC outputs of the homodyne detectors, we can determine in which channel
the error is occurring (see Table 1). If a syndrome mode does not contain
the error in a certain channel, the DC output of the corresponding detector
will be a straight line without any fluctuation. When the error appearing in
a syndrome mode, the DC output of the corresponding detector will be a line
with fluctuation (coming from the error). A four-channel digital
oscilloscope is used to record the DC output of detectors D$_{1}$-D$_{4}$.
Figure 2 shows error syndrome measurement results. In Figure 2(a), outputs
with fluctuation are obtained by detectors D$_{1}$, D$_{2}$ and D$_{3}$, and
the fluctuations of detectors D$_{1}$ and D$_{3}$ are in-phase. The output
of D$_{4}$ is a straight line because the syndrome mode $\hat{d}_{4}$ does
not contain the error in channel 1 ($\hat{e}_{1}$).\ Comparing this result
with table 1, we can identify that an error is occurring in channel 1. In
Figure 2(b), we have outputs with fluctuation for detectors D$_{1}$, D$_{2}$
and D$_{3}$, and the outputs of detectors D$_{1}$ and D$_{3}$ are
out-of-phase, which means that an error is occurring in channel 2. With the
same way, we know that the error occurs in channels 3, 4 and 5 from the
measured results in Figure 2(c), 2(d) and 2(e), respectively.

\begin{table}[tbp]
\begin{tabular}{lll}
\multicolumn{3}{l}{\textbf{Table 1 Error syndrome measurements.}} \\ \hline
The error & Detectors with \ \ \ \ \ \ \ \ \ \ \ \ \  & Measurement \\ 
channel \  & fluctuation & basis \\ \hline
1 & 1, 3 (in-phase) & x \\ 
& 2 & p \\ 
2 & 1, 3 (out-of-phase) & x \\ 
& 2 & p \\ 
3 & 3 & x \\ 
& 2 & p \\ 
4 & 3, 4 (out-of-phase) & x \\ 
& 2 & p \\ 
5 & 3, 4 (in-phase) & x \\ 
& 2 & p \\ \hline
\end{tabular}%
\end{table}

\subsection{Error-correction.}

After the position of the error is identified, we can correct the error by
feedfowarding the measurement results of the corresponding homodyne
detectors D$_{1}$-D$_{4}$ to the output state with suitable gains (see Table
2). The partial encoding method simplifies the error correction procedure.
When the error is occurring in channels 1 and 2, we do not need to correct
it because it does not affect the output state. When the error occurs in the
channel 3, 4 or 5, the output state will be stained by the error and we need
to implement the feedforward of the measurement results.


Figure 3 shows the results of QEC procedure for a vacuum input. The
correction results for an error occurring in channels 1-5 are shown in
Figure 3(a)-3(e), respectively. The quadrature components of output states
before the error correction (cyan line), and after the correction (red and
blue line) are given, where the red and blue lines correspond to the case
using the squeezed and coherent state to be the ancilla modes, respectively,
the black lines are the SNL. From Figure 3(a) and 3(b), we can see that the
output state is immune from errors appearing in channels 1 and 2. Thus, we
do not need to perform error correction when errors are occurring in
channels 1 and 2. When the error is imposed on channels 3, 4 and 5, the
output state contains the error signal before the error correction [cyan
lines in Figure 3(c), 3(d) and 3(e)]. In the error correction procedure, the
measurement results of detectors 3 (or 4) and 2 are fedforward to the output
state (see Table 2). Figure 3(c)-3(e) show, when the squeezed ancilla modes
are utilized, the noises on the output state are reduced. The better the
squeezing, the lower the noise of output state. When the used ancilla modes
are perfect squeezed states, the output state will totally overlap with the
input vacuum state. The measured noise power of the output state can be
found in APPENDIX C.

\begin{figure}[tbp]
\begin{center}
\includegraphics[width=80mm]{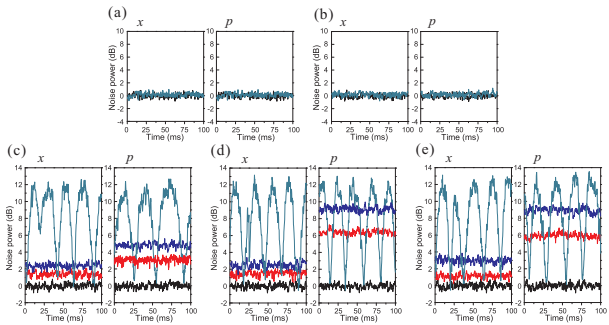}
\end{center}
\caption{(color online) The error correction results for a vacuum input.
(a)-(e) are the results of error correction with an error on channel 1-5,
respectively. Black lines: the SNL. Cyan lines: the noises on amplitude (x)
and phase (p) components of output state before error correction. Blue and
red lines are the noises on x and p components of output state with the
coherent and squeezed ancilla modes, respectively. Measurement frequency is
2 MHz, the spectrum analyzer resolution bandwidth is 30 kHz, and the video
bandwidth is 300 Hz.}
\end{figure}

QEC results with a phase-squeezed state ($-3.5$ dB $/$ $8.9$ dB
squeezing/antisqueezing) as the input state are shown in Figure 4. Figure
4(a)-4(e) are the results of the corrections for an error in channels 1-5,
respectively. In Figure 4(a) and 4(b), the output state is still a phase
squeezed state before the error correction (cyan line) when errors are
occurring in channels 1 and 2, which shows that the output state is not
affected by errors in channels 1 and 2. The measured squeezing and
antisqueezing of the output state are $-2.78$ dB $/$ $8.22$\ dB and $-2.73$
dB $/$ $8.09$\ dB for the errors in channels 1 and 2, respectively. The
decrease of the squeezing derives from the imperfection in the experiment,
such as channel loss and fluctuation of phase locking system. When the error
is imposed on channel 3, 4 and 5, the output state becomes very noisy before
error correction (cyan line). After error correction, the measured noise of
the output state with the squeezed ancilla modes (red line) is below that
using\ coherent states as the ancilla modes (blue line).

\begin{figure}[tbp]
\begin{center}
\includegraphics[width=80mm]{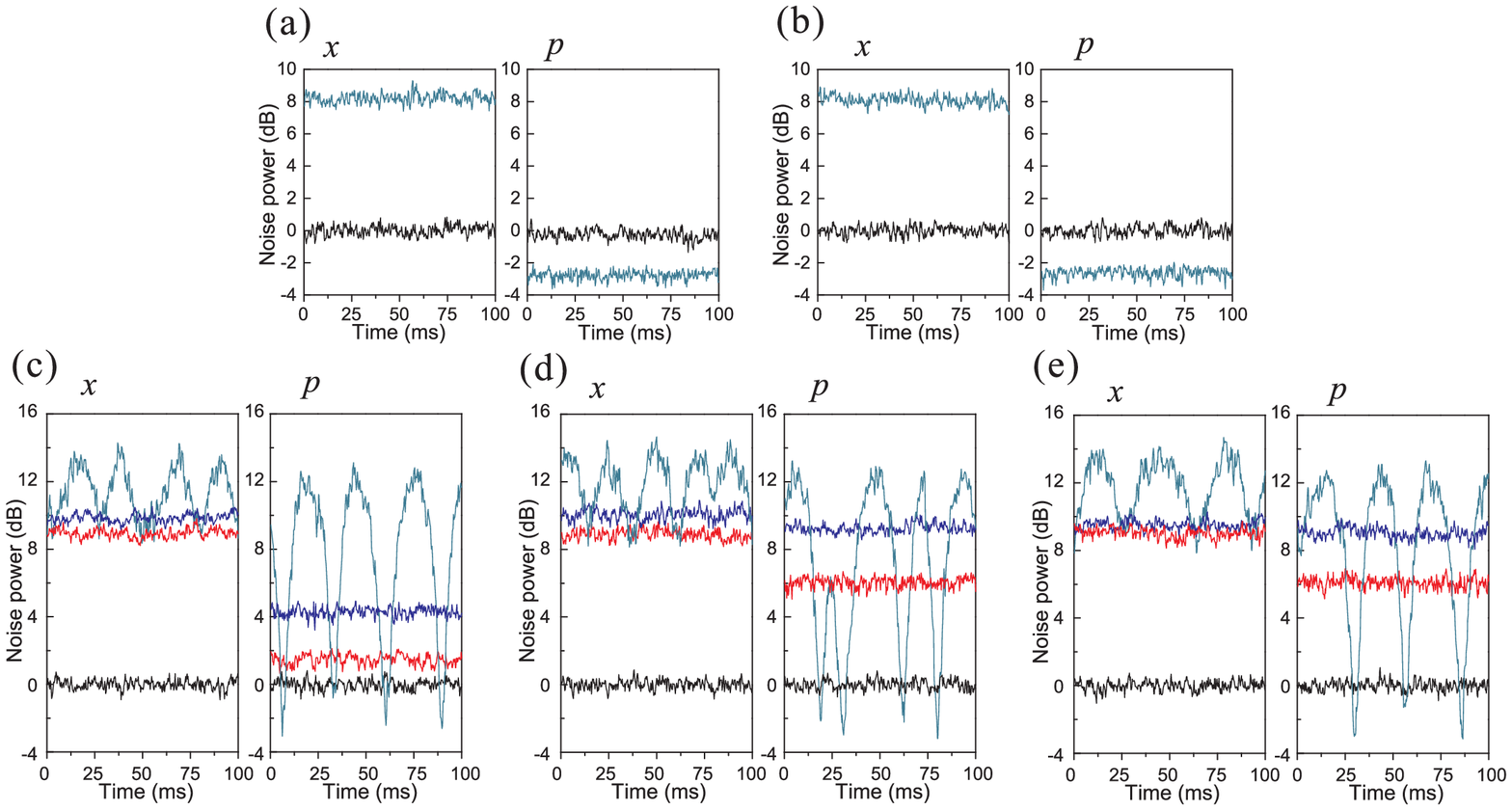}
\end{center}
\caption{(color online) The error correction results for a phase-squeezed
input. (a)-(e) are the results of error correction with an error on channel
1-5, respectively. Black lines: the SNL. Cyan lines: the noises of the
amplitude (x) and phase (p) components of output state before the error
correction. Blue and red lines correspond to the noises levels of output
state after the error correction with the coherent and squeezed ancilla
modes, respectively. Measurement frequency is 2 MHz, the spectrum analyzer
resolution bandwidth is 30 kHz, and the video bandwidth is 300 Hz.}
\end{figure}

The fidelity $F=\left\{ \text{Tr}[(\sqrt{\hat{\rho}_{1}}\hat{\rho}_{2}\sqrt{%
\hat{\rho}_{1}})^{1/2}]\right\} ^{2}$, which denotes the overlap between the
experimentally obtained output state $\hat{\rho}_{2}$\ and the input state $%
\hat{\rho}_{1}$, is utilized to quantify the performance of the QEC code.
The fidelity for two Gaussian states $\hat{\rho}_{1}$\ and $\hat{\rho}_{2}$\
with the covariance matrices $\sigma _{j}$\ is expressed by \cite%
{Nha2005,Scutaru1998} 
\begin{equation}
F=\frac{2}{\sqrt{\Delta +\sigma }-\sqrt{\sigma }}\exp [-\mathbf{\beta }^{T}(%
\mathbf{\sigma }_{1}+\mathbf{\sigma }_{2})^{-1}\mathbf{\beta }],
\end{equation}%
where $\Delta =\det (\sigma _{1}+\sigma _{2}),$\ $\sigma =(\det \sigma
_{1}-1)(\det \sigma _{2}-1),$\ $\beta =\alpha _{2}-\alpha _{1},$and $\alpha
_{j}$\ is the mean amplitudes $\alpha _{j}\equiv (\alpha _{jx},\alpha
_{jp})^{T}$\ ($j=1,2$), $\sigma _{1}$\ and $\sigma _{2}$\ are the covariance
matrices for the input state ($\hat{\rho}_{1}$) and the experimentally
obtained output state ($\hat{\rho}_{2}$), respectively. In our experiment, a
vacuum state and a squeezed vacuum state are used for the input states,
respectively, and the mean amplitude for the both states equals to zero. If
squeezed states with infinite squeezing ($r\rightarrow \infty $) are
utilized as the ancilla states, the fidelity will equal to 1. When\ all\
ancilla\ modes\ are\ the coherent\ states\ of light with zero classical
noise ($r=0$),\ the\ obtained\ fidelity\ of\ the\ output\ state\ is\ the\
corresponding\ classical\ limit \cite{Aoki2009,Lassen2010}.\textbf{\ }Since
the errors in channels 1 and 2 do not affect the output state, the obtained
fidelity is near unity (0.99). The fidelity obtained with squeezed states to
be the ancilla modes is higher than that obtained with coherent states when
error appears in channel 3, 4 and 5 (see Table 2).

\begin{table}[tbp]
\begin{tabular}{lllll}
\multicolumn{5}{l}{\textbf{Table 2 Error correction feedforward components }}
\\ 
\multicolumn{5}{l}{\textbf{and the obtained fidelities.}} \\ \hline
Error \ \  & Quadra- & Feedforward \  & Fidelity \ \ \ \  & Fidelity \ \ \ \ 
\\ 
in & ture of & components \  & with cohe- \  & with \\ 
channel & output &  & rent state \  & squeezing \\ \hline
1 & x & 0 & 0.99 (0.99) & 0.99 (0.99) \\ 
& p & 0 &  &  \\ 
2 & x & 0 & 0.99 (0.99) & 0.99 (0.99) \\ 
& p & 0 &  &  \\ 
3 & x & $\sqrt{2/3}x_{_{3}}$ & 0.60 (0.68) & 0.75 (0.85) \\ 
& p & $-\sqrt{2}p_{_{2}}$ &  &  \\ 
4 & x & $\sqrt{2/3}x_{_{4}}$ & 0.40 (0.42) & 0.56 (0.60) \\ 
& p & $2\sqrt{2}p_{_{2}}$ &  &  \\ 
5 & x & $-\sqrt{2/3}x_{_{4}}$ & 0.39 (0.44) & 0.59 (0.59) \\ 
& p & $2\sqrt{2}p_{_{2}}$ &  &  \\ \hline
\multicolumn{5}{l}{Fidelities in and out of brackets are for the case of a
squeezed} \\ 
\multicolumn{5}{l}{and a vacuum state used as input state, respectively.} \\ 
&  &  &  & 
\end{tabular}%
\end{table}

\section{Discussion}

The presented compact five-wave-packet QEC code can be applied to correct a
single stochastic error in a single quantum channel. For this type of error
correction one usually assume that errors occur stochastically with a small
probability so that multiple errors are unlikely to happen. When two or more
errors are occurring simultaneously on the encoded channels, the errors can
not be identified and corrected because the syndrome measurement will be
confusing \cite{Aoki2009,Lassen2010}.

The general error $\hat{e}=\hat{x}+i\hat{p}$\ ($\hat{x}\neq 0,\quad \hat{p}%
\neq 0$) and $x$-displacement error $\hat{e}=\hat{x}$\ can be well
recognized and corrected suitably with the presented scheme. For the $p$%
-displacement error $\hat{e}=i\hat{p}$, it is unclear which channel the
error comes from since only\emph{\ }the phase measurement of detector D$_{2}$
has output with fluctuation for all five channels (see Table 1). If this
happens in the syndrome measurement results, we need to apply a Fourier
transformation $F$ (a 90$^{\circ }$\ rotation in the phase space) on each
ancilla mode in the encoding stage. In this way, the output state is given
by $U^{-1}[U(F\hat{a}_{1},F\hat{a}_{2},F\hat{a}_{3},\hat{a}_{in},F\hat{a}%
_{4})^{T}+(\hat{e}_{1},\hat{e}_{2},\hat{e}_{3},\hat{e}_{4},\hat{e}%
_{5})^{T}]=(F\hat{a}_{1},F\hat{a}_{2},F\hat{a}_{3},\hat{a}_{in},F\hat{a}%
_{4})^{T}+U^{-1}(\hat{e}_{1},\hat{e}_{2},\hat{e}_{3},\hat{e}_{4},\hat{e}%
_{5})^{T}$ and thus in the syndrome stage, the amplitude quadrature of
detector D$_{2}$ and phase quadratures of detectors D$_{1}$, D$_{3}$, D$_{4}$
are measured. Such that, the $p$-displacement error can be identified by the
outputs with fluctuation from detectors D$_{1}$, D$_{3}$ and D$_{4}$.

In summary, we experimentally demonstrated a compact five-wave-packet CV QEC
code using a five-partite cluster entangled state of light. The QEC code is
implemented only with linear optics operations and four ancilla squeezed
states of light. The compact optics circuit can increase data rates and
decrease chance of further error occurring. The presented partial encoding
method may simplify the error correction procedure and improve the
efficiency of QEC. The presented experiment is the first experimental
demonstration of the approach proposed by S. L. Braunstein and T.
A. Walker for designing linear optics circuits of CV QEC code, which has
potential application in constructing future CV quantum information networks.

\section*{Acknowledgements}

This research was supported by the National Basic Research Program of China
(Grant No. 2010CB923103), NSFC (Grant Nos. 11174188, 61475092) and OIT.

\section*{APPENDIX}

\subsection{Experimental details}

The amplitude-squeezed and phase-squeezed states are produced by three
non-degenerate optical parametric amplifiers (NOPAs) with identical
configuration. These NOPAs are pumped by a common laser source, which is a
continuous wave intracavity frequency-doubled and frequency-stabilized
Nd:YAP/LBO(Nd-doped YAlO$_{3}$ perorskite/lithium triborate) laser \cite%
{WangIEEE2010}. Each of NOPAs consists of an $\alpha $-cut type-II KTP
crystal and a concave mirror \cite{Wang20102}. The front face of the KTP is
coated to be used for the input coupler and the concave mirror serves as the
output coupler of the squeezed states. The transmissions of the input
coupler at 540 nm and 1080 nm are $99.8\%$ and $0.04\%$, respectively. The
transmissions of the output coupler at 540 nm and 1080 nm are $0.5\%$ and $%
5.2\%$, respectively. An NOPA simultaneously generates an amplitude-squeezed
state and a phase-squeezed state in two orthogonal polarizations \cite%
{Yun2000}. The ancilla modes $\hat{a}_{1}$, $\hat{a}_{2}$ and $\hat{a}_{3}$, 
$\hat{a}_{4}$ and the phase-squeezed input state, are generated by three
NOPAs respectively. Three NOPAs are locked individually by using
Pound-Drever-Hall method with a phase modulation of 56 MHz on 1080 nm laser
beam. All NOPAs are operated at deamplification condition, which corresponds
to lock the relative phase between the pump laser and the injected signal to
(2n+1)$\pi $ (n is the integer).

The transmission efficiency of an optical beam from NOPA to a homodyne
detector is around 96\%. The quantum efficiency of a photodiode
(FD500W-1064, Fermionics) used in the homodyne detection system is 95\%. The
interference efficiency on a beam-splitter is about 99\%.

The Fourier transformation $F$ needed for the correction of $p$-displacement
error is a 90$^{\circ }$\ rotation in the phase space, which changes the
squeezing direction of the squeezed state. The Fourier transformations on
the ancilla modes $\hat{a}_{1}$ and $\hat{a}_{4}$ can be completed by
changing the relative phase difference on the beam-splitters T$_{3}$ and T$%
_{4}$ from 0 to $\pi /2$, respectively. The Fourier transformations on the
ancilla modes $\hat{a}_{2}$ and $\hat{a}_{3}$ can be implemented by
exchanging the position of $\hat{a}_{2}$ and $\hat{a}_{3}$ on the
beam-splitter T$_{1}$, which can be simply achieved by rotating the half
wave-plate for 45$^{\circ }$ placed at the output port of the NOPA that is
because $\hat{a}_{2}$ and $\hat{a}_{3}$ are produced from one NOPA \cite%
{Yun2000}.

\subsection{Details of syndrome and error-correction procedure}

When the error is occurring in channel 1, we have non-zero syndrome
measurement for 
\begin{eqnarray}
\hat{x}_{_{D1}} &=&\frac{\hat{x}_{e_{1}}}{\sqrt{2}}+\hat{x}_{1}^{(0)}e^{-r},
\notag \\
\hat{x}_{_{D3}} &=&\frac{\hat{x}_{e_{1}}}{2\sqrt{2}}+\hat{x}_{3}^{(0)}e^{-r},
\notag \\
\hat{p}_{_{D2}} &=&\frac{3\hat{p}_{e_{1}}}{2\sqrt{6}}+\hat{p}%
_{2}^{(0)}e^{-r},
\end{eqnarray}%
where the outputs of $\hat{x}_{_{D1}}$ and $\hat{x}_{_{D3}}$ are in-phase.
At this case, the output state is 
\begin{eqnarray}
\hat{x}_{out} &=&\hat{x}_{in},  \notag \\
\hat{p}_{out} &=&\hat{p}_{in},
\end{eqnarray}%
which is immune from the error in channel 1, thus we do not need any
correction.

When the error is occurring in channel 2, non-zero syndrome measurement is
obtained for 
\begin{eqnarray}
\hat{x}_{_{D1}} &=&\frac{\hat{x}_{e_{2}}}{\sqrt{2}}+\hat{x}_{1}^{(0)}e^{-r},
\notag \\
\hat{x}_{_{D3}} &=&\frac{-\hat{x}_{e_{2}}}{2\sqrt{2}}+\hat{x}%
_{3}^{(0)}e^{-r},  \notag \\
\hat{p}_{_{D2}} &=&\frac{-3\hat{p}_{e_{2}}}{2\sqrt{6}}+\hat{p}%
_{2}^{(0)}e^{-r},
\end{eqnarray}%
where $\hat{x}_{_{D1}}$ and $\hat{x}_{_{D3}}$ are out-of-phase. The
corresponding output state is%
\begin{eqnarray}
\hat{x}_{out} &=&\hat{x}_{in},  \notag \\
\hat{p}_{out} &=&\hat{p}_{in},
\end{eqnarray}%
and we do not need any correction.

When the error is occurring in channel 3, non-zero syndrome measurements are
obtained for 
\begin{eqnarray}
\hat{x}_{_{D3}} &=&\frac{-2\hat{x}_{e3}}{2\sqrt{2}}+\hat{x}_{3}^{(0)}e^{-r},
\notag \\
\hat{p}_{_{D2}} &=&\frac{2\hat{p}_{e3}}{2\sqrt{6}}+\hat{p}_{2}^{(0)}e^{-r}.
\end{eqnarray}%
In this case, the output state is%
\begin{eqnarray}
\hat{x}_{out} &=&\hat{x}_{in}+\frac{\hat{x}_{e_{3}}}{\sqrt{3}},  \notag \\
\hat{p}_{out} &=&\hat{p}_{in}+\frac{\hat{p}_{e_{3}}}{\sqrt{3}}.
\end{eqnarray}%
To eliminate the error, $\frac{\sqrt{2}}{\sqrt{3}}\hat{x}_{_{D3}}$ and $-%
\sqrt{2}\hat{p}_{_{D2}}$ should be fedforward to $\hat{x}_{out}$ and $\hat{p}%
_{out}$, respectively. The corrected output mode is given by 
\begin{eqnarray}
\hat{x}_{out}^{\prime } &=&\hat{x}_{in}+\frac{\sqrt{2}}{\sqrt{3}}\hat{x}%
_{3}^{(0)}e^{-r},  \notag \\
\hat{p}_{out}^{\prime } &=&\hat{p}_{in}-\sqrt{2}\hat{p}_{2}^{(0)}e^{-r}.
\end{eqnarray}%
The noise powers of the output state are 
\begin{equation}
\left\langle \Delta ^{2}\hat{x}_{out}^{\prime }\right\rangle =\left\langle
\Delta ^{2}\hat{x}_{in}\right\rangle +\frac{2}{3}\times \frac{1}{4}e^{-2r}
\end{equation}%
and 
\begin{equation}
\left\langle \Delta ^{2}\hat{p}_{out}^{\prime }\right\rangle =\left\langle
\Delta ^{2}\hat{p}_{in}\right\rangle +2\times \frac{1}{4}e^{-2r},
\end{equation}%
respectively.

When the error is occurring in channel 4, we have non-zero syndrome
measurements on 
\begin{eqnarray}
\hat{x}_{_{D3}} &=&\frac{-\hat{x}_{e_{4}}}{2\sqrt{2}}+\hat{x}%
_{3}^{(0)}e^{-r},  \notag \\
\hat{x}_{_{D4}} &=&\frac{\hat{x}_{e_{4}}}{\sqrt{2}}+\hat{x}_{4}^{(0)}e^{-r},
\notag \\
\hat{p}_{_{D2}} &=&\frac{\hat{p}_{e_{4}}}{2\sqrt{6}}+\hat{p}_{2}^{(0)}e^{-r},
\end{eqnarray}%
where $\hat{x}_{_{D3}}$ and $\hat{x}_{_{D4}}$ are out-of-phase. The
corresponding output state is 
\begin{eqnarray}
\hat{x}_{out} &=&\hat{x}_{in}-\frac{\hat{x}_{e4}}{\sqrt{3}},  \notag \\
\hat{p}_{out} &=&\hat{p}_{in}-\frac{\hat{p}_{e4}}{\sqrt{3}}.
\end{eqnarray}%
The measurement results of $\frac{\sqrt{2}}{\sqrt{3}}\hat{x}_{_{D4}}$ and $2%
\sqrt{2}\hat{p}_{_{D2}}$ should be fedforward to $\hat{x}_{out}$ and $\hat{p}%
_{out}$ to eliminate the error. The corrected output mode is 
\begin{eqnarray}
\hat{x}_{out}^{\prime } &=&\hat{x}_{in}+\frac{\sqrt{2}}{\sqrt{3}}\hat{x}%
_{4}^{(0)}e^{-r},  \notag \\
\hat{p}_{out}^{\prime } &=&\hat{p}_{in}+2\sqrt{2}\hat{p}_{2}^{(0)}e^{-r},
\end{eqnarray}%
and the corresponding noise powers are 
\begin{equation}
\left\langle \Delta ^{2}\hat{x}_{out}^{\prime }\right\rangle =\left\langle
\Delta ^{2}\hat{x}_{in}\right\rangle +\frac{2}{3}\times \frac{1}{4}e^{-2r},
\end{equation}%
and%
\begin{equation}
\left\langle \Delta ^{2}\hat{p}_{out}^{\prime }\right\rangle =\left\langle
\Delta ^{2}\hat{p}_{in}\right\rangle +8\times \frac{1}{4}e^{-2r},
\end{equation}%
respectively.

When the error is occurring in channel 5, we have non-zero syndrome
measurements on 
\begin{eqnarray}
\hat{x}_{_{D3}} &=&\frac{\hat{x}_{e5}}{2\sqrt{2}}+\hat{x}_{3}^{(0)}e^{-r}, 
\notag \\
\hat{x}_{_{D4}} &=&\frac{\hat{x}_{e_{5}}}{\sqrt{2}}+\hat{x}_{4}^{(0)}e^{-r},
\notag \\
\hat{p}_{_{D2}} &=&\frac{-\hat{p}_{e5}}{2\sqrt{6}}+\hat{p}_{2}^{(0)}e^{-r},
\end{eqnarray}%
where $\hat{x}_{_{D3}}$ and $\hat{x}_{_{D4}}$ are in-phase. The output state
is 
\begin{eqnarray}
\hat{x}_{out} &=&\hat{x}_{in}+\frac{\hat{x}_{e_{5}}}{\sqrt{3}},  \notag \\
\hat{p}_{out} &=&\hat{p}_{in}+\frac{\hat{p}_{e_{5}}}{\sqrt{3}}.
\end{eqnarray}%
By feedforwarding $\frac{-\sqrt{2}}{\sqrt{3}}\hat{x}_{_{D4}}$ and $2\sqrt{2}%
\hat{p}_{_{D2}}$ to $\hat{x}_{out}$ and $\hat{p}_{out}$, the error will be
corrected. The output mode after correction is 
\begin{eqnarray}
\hat{x}_{out}^{\prime } &=&\hat{x}_{in}-\frac{\sqrt{2}}{\sqrt{3}}\hat{x}%
_{4}^{(0)}e^{-r},  \notag \\
\hat{p}_{out}^{\prime } &=&\hat{p}_{in}+2\sqrt{2}\hat{p}_{2}^{(0)}e^{-r}
\end{eqnarray}%
and the corresponding noise powers are 
\begin{equation}
\left\langle \Delta ^{2}\hat{x}_{out}^{\prime }\right\rangle =\left\langle
\Delta ^{2}\hat{x}_{in}\right\rangle +\frac{2}{3}\times \frac{1}{4}e^{-2r},
\end{equation}%
and%
\begin{equation}
\left\langle \Delta ^{2}\hat{p}_{out}^{\prime }\right\rangle =\left\langle
\Delta ^{2}\hat{p}_{in}\right\rangle +8\times \frac{1}{4}e^{-2r},
\end{equation}%
respectively.

\begin{table*}[tbp]
\begin{tabular}{llll}
\multicolumn{4}{l}{\textbf{Table C1 The noise powers of the output state
(with the unit of dB). }} \\ \hline
Error \ \  & Quadra- \ \  & Noise of the output state \ \  & Noise of the
output state \ \  \\ 
in & ture of & without squeezing & with squeezing \\ 
channel & output & on ancilla modes\ \ \ \ \  & on ancilla modes \ \ \ \ \ \ 
\\ \hline
1 & x & $0.15\pm 0.30$ ($8.22\pm 0.31$) &  \\ 
& p & $0.13\pm 0.30$ ($-2.78\pm 0.27$) &  \\ 
2 & x & $0.19\pm 0.29$ ($8.09\pm 0.31$) &  \\ 
& p & $0.18\pm 0.30$ ($-2.73\pm 0.29$) &  \\ 
3 & x & $2.39\pm 0.28$ ($9.85\pm 0.27$) & $1.37\pm 0.29$ ($8.93\pm 0.27$) \\ 
& p & $4.80\pm 0.29$ ($4.28\pm 0.28$) & $3.07\pm 0.31$ ($1.46\pm 0.30$) \\ 
4 & x & $2.47\pm 0.34$ ($9.96\pm 0.32$) & $1.49\pm 0.29$ ($8.89\pm 0.29$) \\ 
& p & $9.13\pm 0.30$ ($9.25\pm 0.27$) & $6.40\pm 0.28$ ($6.04\pm 0.30$) \\ 
5 & x & $2.99\pm 0.31$ ($9.51\pm 0.27$) & $1.14\pm 0.28$ ($9.02\pm 0.30$) \\ 
& p & $9.01\pm 0.32$ ($9.03\pm 0.32$) & $5.94\pm 0.30$ ($6.10\pm 0.33$) \\ 
\hline
\multicolumn{4}{l}{The noise powers of the output state in and out of
brackets are for the case} \\ 
\multicolumn{4}{l}{of a squeezed and a vacuum state used as input state,
respectively.}%
\end{tabular}%
\end{table*}

\subsection{Noise power of the output state}

The measured noise power of the output state in QEC is shown in table C1.
Measurement frequency of noise power is 2 MHz, the spectrum analyzer
resolution bandwidth is 30 kHz, and the video bandwidth is 300 Hz.

\end{document}